\documentclass[pra,twocolumn,showpacs,superscriptaddress]{revtex4-1}
\usepackage{amsmath,amscd,amsfonts,amssymb,color,bm}
\usepackage{graphicx,amsfonts,epsf,multirow}
\usepackage{float,verbatim}
\usepackage{braket}
\usepackage{dsfont}

\begin{document}

\title{Role of Bell-CHSH violation and local filtering in quantum key distribution}
\author{Jaskaran Singh}
\email{jaskaransinghnirankari@iisermohali.ac.in}
\affiliation{Department of Physical Sciences,
Indian
Institute of Science Education \&
Research (IISER) Mohali, Sector 81 SAS Nagar,
Manauli PO 140306 Punjab India.}

\author{Sibasish Ghosh}
\email{sibasish@imsc.res.in}
\affiliation{Optics \& Quantum Information Group,
Institute of Mathematical Sciences, HBNI, C.I.T Campus, Taramani,
Chennai 600113, India}

\author{Arvind}
\email{arvind@iisermohali.ac.in}
\affiliation{Department of Physical Sciences,
Indian
Institute of Science Education \&
Research (IISER) Mohali, Sector 81 SAS Nagar,
Manauli PO 140306 Punjab India.}

\author{Sandeep K.~Goyal}
\email{skgoyal@iisermohali.ac.in}
\affiliation{Department of Physical Sciences,
Indian
Institute of Science Education \&
Research (IISER) Mohali, Sector 81 SAS Nagar,
Manauli PO 140306 Punjab India.}
\begin{abstract}
In this article, we analyse the relationship between the Bell violation and the secure key rate of entanglement assisted quantum key distribution (QKD) protocols. Specifically, we address the question whether Bell violation is necessary or sufficient for secure communication.  We construct a class of states which do not show Bell violation, however, which can be used for secure communication after local filtering. Similarly, we identify another class of states which show Bell violation but can not be used for generating secure key even after local filtering.  The existence of these two classes of states demonstrates that Bell violation as an initial resource is neither necessary nor sufficient for QKD,
while it becomes necessary after filteration. Our work therefore forces a departure from traditional thinking that the degree of Bell violation is a key resource for quantum communication and brings out the role of local filtering.
\end{abstract}

\maketitle

\section{Introduction}
Quantum correlations have been instrumental in the development of quantum key distribution (QKD) protocols, where two parties Alice and Bob establish a secret key for secure communication \cite{Crypt_review, Security_review, BB84,chandan, jaskaran}. QKD protocols can be classified into two different classes. The first class contains the prepare and measure schemes which involve Alice preparing the system in one of many possible states and transmitting it to Bob. Bob then performs a measurement on the same. Afterwards both the parties perform basis reconciliation and distill out a secret key. Examples of such schemes include BB84 \cite{BB84}, B92 \cite{B92}, six state protocol \cite{Six_state} and SARG04 \cite{SARG_2004}. Another popular protocol under prepare and measure scheme is the measurement device independent QKD protocol~\cite{Yin2016,Tang2014,Lo2012,Pirandola2015,Liu2013,Tang2014a}. In this protocol, both Alice and Bob prepare their systems in states of their choice and send them to a third party Charlie, who performs joint measurements on these systems. Afterwards Charlie discloses his measurement results, the knowledge of which Bob (or Alice) uses to  apply a unitary operation to his state which is then used to generate a secure key. The second class of QKD schemes involving the use of entanglement shared between Alice and Bob~\cite{Entanglement} and are termed as entanglement assisted QKD protocols, e.g. the E91 protocol \cite{E91}. Recently, entanglement based versions of measurement device independent protocols have also been studied~\cite{Xu2013,Yang2016}. Protocols involving two way communication between the parties have been shown to allow higher error rates than 
standard one way communication schemes and have been studied 
for both prepare and measure~\cite{lo_two_way_PM} and entanglement based QKD scheme~\cite{two_way_EB,renner_two_way_EB, Acin_sec}. 
 
 QKD Protocols in both classes are proven to be robust against eavesdropping \cite{Shor_Pres_2000, Perform_BB84, Sec_BB84, Acin_sec, Sec_monogamy} and are fundamentally secure as opposed to the classical key distribution protocols.

Although, the security of the entanglement based QKD protocol can be proven by comparing the information content of the eavesdropper to the information content of the two involved parties~\cite{lutkenhaus_key}, violation of a Bell type inequality is necessary for the security of the protocol~\cite{acin_privacy, Brukner_sec}.   It is also known that entanglement is necessary but not sufficient to violate a Bell's inequality~\cite{Werner}.  This renders a huge class of entangled states unusable for entanglement based QKD. On the  other hand, the question whether Bell violation is sufficient for the security of  QKD is also not settled and has been a matter of debate~\cite{Acin_sec}.  Since
Bell-CHSH violation requires entanglement, it is important to ask if one can carry out 
QKD schemes with states which are entangled but do not violate Bell's inequalities. 

 Any Bell type inequality~\cite{EPR,Nonlocality} such as the CHSH inequality~\cite{CHSH}, I$_{3322}$ inequality~\cite{Froissart_1981, Collins_2004}, as well as the CGLMP inequality~\cite{CGLMP}, characterizes the non-classicality of the correlations. Due to the lack of analytical results of Bell violation  for more than two measurement on each parties, we consider only the CHSH inequality in this article. However, it is worthwhile to mention that  states that do not violate CHSH inequality may violate some other Bell inequality with  higher number of measurement settings. A particular example for the same  can be found in Ref.~\cite{Collins_2004}.

In this article, we propose a geometrical representation of correlations and relate the CHSH violation with the secure key rate for the protocols where the secure key rate is a function of the error rate only. Such a representation allows direct inference of secure key rate alongwith Bell-CHSH violation, and states offering optimal security can be directly identified. Using this representation we identify a class of states showing Bell violation but not offering a secure key. These states are therefore of no use for QKD as they lead to a higher error rate and no security. Further, we show that using local filtering operations some states which initially showed no Bell-CHSH violation can be made to provide non-zero secure key rate.  Hence, we conclusively prove that Bell-CHSH violation as an initial resource is neither  necessary nor sufficient for the security of QKD protocols, thereby answering an open question put forward by Acin \textit{et.~al.}~\cite{Acin_sec}. Furthermore, we propose a modified QKD protocol which uses local filtering to acquire higher secure key rate.

We derive our results based only on two assumptions: i) We
consider those entanglement based protocols for which the secure key rate is a function of error rate only, and ii) Bell-CHSH
violation is necessary in order to ensure that the correlations shared by the parties can 
give rise to a secure key. The protocols we consider
also encompass one way or two way communication schemes for entanglement based QKD and our results are
therefore more general.

Classical post processing  schemes like advantage distillation has been previously considered in the literature to improve the security of QKD protocols~\cite{acin_classical_distillation}. It has also been shown that there exist bipartite bound entangled states which initially do not violate any known Bell's inequalities, but can be transformed using local operations and classical communications to states from which  a secure key can be distilled~\cite{horodecki_bound_secure_key}. However, these states still do not violate any known Bell's 
inequality. There are states that violate Bell-CHSH inequality and still cannot be transformed into states useful for QKD. Since these states are entangled, under general multicopy entanglement distillation these states can in principle be made useful
for QKD. 

Here we use local quantum filtering to alter the secure key rate and Bell violation. Local filtering operations allow states to concentrate entanglement~\cite{debmalya} and may reveal hidden Bell non-locality \cite{Filtering} and can therefore increase the secure key rate.
Local filtering is  a special class of entanglement distillation and can be applied to single copies.  Moreover, since  QKD protocols are generally implemented on photonic systems, single copy operations are more practical than multicopy operations. Therefore, experimentally local filtering is much more accessible than multicopy entanglement distillation~\cite{filtering_expt}.

The article is organised as follows: In Section~\ref{sec:
back} we review the Bell-CHSH inequality, outline a general
entanglement assisted QKD protocol and give a brief
description of local filtering operations. In
Section~\ref{sec:res} we develop a geometrical
representation of correlations which provides a clear
picture of how various states would fare for QKD and show
that application of local filtering operations is indeed
advantageous. In Section~\ref{sec: conc} we offer concluding
remarks and discussions.

\section{Background}
\label{sec: back}
In this section we provide the relevant background
with an aim 
to 
calculate various quantities such as the Bell-CHSH violation
and the secure key rate in entanglement assisted QKD
protocols.
In
Section.~\ref{sec:filtering} we briefly outline local
filtering operations on two-qubit systems.

\subsection{Bell-CHSH inequality}
\label{sec: bell}
The Bell-CHSH inequality quantifies the correlations arising from local measurements on two-qubit states. All correlations which violate the inequality are termed as non-local as they defy explanation by any local realistic hidden variable model (LRHVM).

The Bell-CHSH inequality involves two parties Alice and Bob sharing an entangled state $\rho$.  Each party performs one of the two measurements, having two outcomes $\pm 1$ on their respective subsystem. Let $A_0,A_1$ be the measurements performed on Alice's particle and $B_0,B_1$ be the measurements performed in Bob's particle. We can define a joint operator $\mathcal{B} = A_0\otimes B_0  +  A_0\otimes B_1 + A_1\otimes B_0  - A_1\otimes B_1 $ which is called the Bell operator. The Bell-CHSH inequality states that the expectation value $S$ of the Bell operator $\mathcal{B}$ for the classical situations describable by LRHVM is bounded between $2$ and $-2$, i.e.,
\begin{equation}
  |S| \equiv |\mathcal{B}| \leq 2. \label{eq: BCI} \end{equation}
However, some quantum states violate this bound implying that there is no LRHVM for the corresponding measurement scenarios.

An arbitrary  two-qubit state $\rho$ can be written in the Hilbert-Schmidt form as~\cite{wilde_2013}
\begin{equation} \rho =
\frac{1}{4}[ \mathds{1}\otimes\mathds{1} +
\bm{r}\cdot\bm{\sigma}\otimes\mathds{1} +
\mathds{1}\otimes\bm{s}\cdot\bm{\sigma} + \sum_{i,j = 1}^3
T_{ij}\sigma_i\otimes\sigma_j], \label{eq: den}
\end{equation} 
where $\bm{r}$ and $\bm{s}$ are three-dimensional real vectors characterizing the reduced density matrices of the first and the second qubit respectively and $T$ is a $3\times 3$  real matrix representing the correlations between the two qubits. The state $\rho$ can also be parameterized linearly with real parameters as
\begin{equation} 
\rho = \frac{1}{4}\sum_{i,j =
0}^3M_{ij}\sigma_i\otimes\sigma_j, 
\label{eq:muellermatrix}
\end{equation} 
where $\sigma_0$ is the $2\times 2$ identity matrix and $M_{ij}$ is the Mueller matrix \cite{simon_1994}, with $M_{00} = \text{Tr}(\rho)$, $M_{0j} = \bm{s}_j$, $M_{i0} = \bm{r}_i$ and $M_{ij} = T_{ij}~\forall i,j \in \lbrace1,2,3\rbrace$. This representation turns out to be quite useful as will become evident.

The measurement operators $\{A_0,A_1\}$ and $\{B_0,B_1\}$ are defined as
\begin{align}
    A_i &= \bm{a}_i\cdot \bm\sigma,\qquad
    B_i = \bm{b}_i\cdot \bm\sigma,
   \end{align}
where $\bm{a}_i$ and $\bm{b}_i$ are normalized three-dimensional real vectors $\forall i\in \lbrace 0, 1\rbrace$. In this new notation, we can calculate the expectation value of the Bell operator as
\begin{equation}
S = \bm{a}_0^t T \bm{b}_0 + \bm{a}_0^t T
\bm{b}_1 + \bm{a}_1^t T \bm{b}_0 - \bm{a}_1^t T \bm{b}_1.
\label{eq: nchsh}
\end{equation}
Simple algebra shows that for a given two-qubit state $\rho$ the maximum value of $S$ that can be achieved for optimal measurements is~\cite{tsirelson}
\begin{align}
\text{max}\lbrace S\rbrace &=  2\sqrt{\lambda_1^2 + \lambda_2^2},	
\label{eq: maxs2}
\end{align}
where $\lambda_1$ and $\lambda_2$ are the two largest singular values of the correlation matrix $T$ each of which is  bounded from above by $1$. Therefore, the maximum Bell-CHSH violation is achieved when $S = 2\sqrt{2}$~\cite{tsirelson2}.

An interesting point to note is that the violation of the Bell-CHSH inequality does not depend on the Bloch vectors $\bm{r}$ and $\bm{s}$, but only on the correlation matrix $T$. Therefore, different states with the same correlation matrix result in the same value of $S$ which itself is determined by the two parameters  $\lambda_1$ and $\lambda_2$ only.

Therefore, if we fix the optimized Bell violation parameter $S$ we obtain a relation between $\lambda_1$ and $\lambda_2$ giving us a way to describe the family of states with this particular value of Bell violation by only one effective parameter.

\subsection{Entanglement assisted QKD protocols}
\label{sec: protocol}
In this article, we are interested only in those entanglement assisted QKD protocols in which the secure key rate is a function of quantum bit error rate (QBER). This assumption also encompasses any $1$-way or 
$2$-way communication scheme.
In this subsection, we define the QBER and the minimum secure key rate. For simplicity, we consider the entanglement based BB84 protocol.

The two parties Alice and Bob who want to establish a secure key, share a bipartite entangled state $\rho$. Each of them have a choice of $L$ number of $d$-outcome  mutually unbiased measurement bases (MUBs), where $d$ is the dimension of each of the subsystem. They perform measurement of the observables randomly chosen from the set of $L$ observables on their respective subsystems and keep a record of the measurement  outcomes. Afterwards, they publicly compare their measurement bases and keep only those outcomes comprising the raw key for which their bases match as the raw key. The parties can then perform information reconciliation and privacy amplification to improve the key which  utilize classical algorithms. Since these algorithms only enhance  the raw key, we do not consider them in our analysis, but rather focus on  the raw key itself.

In the ideal scenario, Alice and Bob are left with perfectly identical keys. However, imperfections in state preparation, transmission and measurement processes can yield differences in their key strings.  Alice and Bob can estimate the QBER $Q$ after comparing a small portion of their secret key. Formally, the QBER $Q$ for a given state $\rho$ is defined as the average mismatch between the outcomes of Alice and Bob. If Alice has $L$ number of MUBs  denoted by $\{\ket{\psi_i^\alpha}\}_{i = 1}^d$ (for $1\le \alpha \le L$) which are correlated to Bob's MUBs $\{\ket{\phi_j^\alpha}\}_{j = 1}^d$, then the perfect correlation between Alice and Bob would imply that whenever Alice and Bob perform measurements in the $\alpha$-th basis and Alice's outcome is $\ket{\psi_i^\alpha}$ then Bob's outcome must be $\ket{\phi_i^\alpha}$. In the non-ideal scenario, there can be non-zero probability of observing $\ket{\psi_i^\alpha}$ in Alice's lab and $\ket{\phi_j^\alpha}$ in Bob's where $i \ne j$. Hence, the QBER which is an average of all these mismatch probabilities can be expressed as:~\cite{lutkenhaus_key}
\begin{equation}
Q = \frac{1}{L}\sum^L_{\alpha=1}\sum_{i\neq j = 1}^d
\langle\psi_i^\alpha\phi_j^\alpha|\rho|\psi_i^\alpha\phi_j^\alpha\rangle.
\label{eq:Error}
\end{equation}

The expression in Eq.~\eqref{eq:Error} for  QBER  holds for any $L\le d+1$ number of MUBs. In this article we restrict our analysis to qubits only, i.e., $d=2$.  For the case of two  measurement basis with each party, QBER can be calculated using 
Eq.~\eqref{eq:Error} as
\begin{align}
  Q&=\frac{1}{4}\left( 2 - \bm{x}_0^t T\bm{y}_0 - \bm{x}_1^t T\bm{y}_1 \right),\nonumber\\
  &\frac{1}{4}\left(2 - |\lambda_1| - |\lambda_2| \right),
\label{eq: error}
\end{align}
where $\bm{x}_i$ and $\bm{y}_j$  are the Bloch  vectors of  the measurement basis with Alice and Bob, respectively, and $T$ is the correlation matrix. $\lambda_1$ and $\lambda_2$ the two largest singular values of the matrix $T$.

Similarly, for the case of $L=3$, in which both the parties have a choice 
of three mutually unbiased measurement basis, QBER can be calculated as,
\begin{equation}
\begin{aligned}
Q &= \frac{1}{6}(3-\bm{x}_0^t
T\bm{y}_0 - \bm{x}_1^t T\bm{y}_1-\bm{x}_2^t T\bm{y}_2)\\
&=\frac{1}{6}(3-|\lambda_1| - |\lambda_2|-|\lambda_3|).
\end{aligned}
\end{equation}

Due to the lack of complete analytical understanding of Bell inequalities using more than two measurement basis, we focus in our work on $L=2$ case which corresponds to the CHSH inequality. 
The minimum secure key rate $r_{min}$ is defined as the average number of secret bits that can be distilled from each run of the protocol when Alice and Bob measured in the same basis. The rate
$r_{min}$ depends on a number of factors including the   strategy incorporated by the eavesdropper. Therefore, there is no general expression for calculating $r_{min}$ for a given value $Q$. In some special cases, one can estimate $r_{min}$ and arrive at an expression. For example, for the case of symmetric attacks by eavesdroppers (as presented in Ref.~\cite{lutkenhaus_key}) in entanglement assisted protocols for qubits with two measurement settings per qubit, the secure key rate is given by~\cite{lutkenhaus_key}
\begin{equation}
  r_{min} =
1 + 2(1-Q)\log_2{(1-Q)} + 2Q\log_2 Q.  \label{eq: securekey}
\end{equation} 
As is evident, minimizing QBER 
maximizes the secure key rate $r_{min}$. Only when $r_{min}>0$, can a secure key be distilled from a protocol. This restricts the QBER to $Q\approx 11\%$ for the symmetric attack. For other protocols it might be more than $11\%$.

It is therefore seen that QBER and the expectation value  $S$ of the Bell operator are   functions of the singular values $\lambda_1,\lambda_2$ of the correlation matrix $T$. 
We have two separate conditions for the security of a QKD protocol. One being $r_{min}>0$ for a secure key to be distilled while the second is the requirement that the underlying  entangled state  violates the CHSH inequality.
While it can be seen that there exist no states with non-vanishing secret key and no Bell-CHSH violation, there 
do exist states which show Bell-CHSH violation but have vanishing secure key.

\subsection{Local filtering}
\label{sec:filtering}
In this subsection, we present a special class of local
quantum operations which
is useful to
concentrate entanglement and non-local correlations in
two-qubit systems.

Local filtering are
operations which transform a state $\rho$ to $\rho'$ which
has a higher concentration of entanglement and Bell
non-local correlations. Consider the local single-qubit measurements on a two-qubit system where the measurement operators $M_1, M_2$ for the first qubit and $N_1, N_2$ for the second qubit. For simplicity, we choose  $M_2 =
\sqrt{\mathds{1} - M_1^\dagger M_1}$ and $N_2 =
\sqrt{\mathds{1} - N_1^\dagger N_1}$.  
The state after measuring $M_1$ and $N_1$ is given by, 
\begin{equation} 
\rho' = \frac{(M_1\otimes
N_1)\rho(M_1\otimes N_1)^\dagger}{\text{Tr}((M_1\otimes
N_1)\rho(M_1\otimes N_1)^\dagger)}.  \label{eq:filtering}
\end{equation}
The entanglement (here the concurrence) in the state $\rho'$ is 
related to the entanglement in $\rho$  as~\cite{conc_wooters, Entanglement},
\begin{equation} C(\rho') =
C(\rho)\frac{|\text{det}(M_1)||\text{det}(N_1)|}{\text{tr}((M_1^\dagger
M_1\otimes N_1^\dagger N_1)\rho)}, \label{Eq:ConcurrenceLF}
\end{equation} and can be made to increase if we consider
operations with $|\text{det}(M_1)|\neq 0$ and
$|\text{det}(N_1)|\neq 0$ and
$|\text{det}(M_1)||\text{det}(N_1)|>\text{Tr}((M_1^\dagger
M_1\otimes N_1^\dagger N_1)\rho)$. It should be noted that
having higher entanglement does not 
necessarily imply higher Bell nonlocality. However, it can be
shown that for a certain class of states, Bell violation can
also be made to increase by employing local filtering~\cite{Gisin1996}. Specifically,
the state $\rho$ can be filtered to a state $\rho'$ which is
Bell diagonal or a special form of the `$X$'
state~\cite{Filtering, simon_1994} and has higher
entanglement and exhibits higher Bell
non-local correlations.

Following Ref.~\cite{Filtering}, we briefly illustrate the
method to obtain a two-qubit filtered state.

Any valid operations on the two qubit state $\rho$ can
be seen as proper orthochronous Lorentz  transformations on
the Mueller matrix $M$ [Eq.~\eqref{eq:muellermatrix}]  as
\begin{equation} 
M' = L_{M_1}ML_{N_1}^T.
\label{eq:lorentztransf} 
\end{equation}

The Lorentz
transformations, $L_{M_1}$ and $L_{N_1}$ are given in terms of the measurement
operators as: 
\begin{equation}
\begin{aligned} L_{M_1} = \frac{V(M_1\otimes
M^*_1)V^\dagger}{|\text{det}(M_1)|}, \\ L_{N_1} =
\frac{V(N_1\otimes N^*_1)V^\dagger}{|\text{det}(N_1)|}, \\
\text{with} ~V=\frac{1}{\sqrt{2}}\begin{pmatrix} 1 & 0 & 0 & 1\\ 0 & 1 &
1 & 0\\ 0 & i & -i & 0\\ 1 & 0 &  0 & -1 \end{pmatrix}.
\end{aligned} 
\end{equation}

Further, the Mueller matrix $M$ can be
brought to a diagonal or a special form
by
Lorentz transformations
$L_1$ and $L_2$ as
\begin{equation}
M = L_1\Sigma L_2^T,
\label{eq:diagonalization}
\end{equation}
where $\Sigma$ is respectively a diagonal Mueller matrix corresponding to
a Bell diagonal state or of the form
\begin{equation}
\Sigma = \begin{pmatrix}
a & 0 & 0 & b\\
0 & d & 0 & 0\\
0 & 0 & -d & 0\\
c & 0 & 0 & a+c-b 
\end{pmatrix},
\end{equation}

where $a, b, c, d$ are real numbers. The latter form can be brought arbitrarily close to
a Bell diagonal state for $d\neq 0$ by application of local filtering operations~\cite{sibasish}, while
$d=0$ corresponds to a separable initial state. A more detailed analysis and 
a geometric picture of these two canonical forms of two qubit states under 
local filtering is given in Ref.~\cite{Filtering,sibasish2}.
A closed form solution relating Bell violation and local filtering can also be found in Ref.~\cite{sibasish}.

The matrix representation of these local filtering
operations applied on the corresponding Mueller matrix can 
be constructed by considering its columns as the eigenvectors of $MGM^TG$ and
its transposition respectively for $L_{M_1}$ and $L_{N_1}$,
where $G = \text{diag}(1,-1,-1,-1)$ is the Minkowski metric. The Mueller matrix $M$ under
these optimal Lorentz transformations 
then transforms as
\begin{equation}
M'=L_1^TGMGL_2.
\end{equation}
Finally, the singular values of the correlation matrix $T$
are the singular values of the matrix $M_{ij}$, 
$i,j\in \lbrace1,2,3\rbrace$ as given in Eq.~\eqref{eq:muellermatrix}.

It is to be noted that the state corresponding to
non-diagonal $\Sigma$ is a
subset of measure zero and thus has zero
probability of
occurrence. Therefore,  most of the
states can be
brought to a Bell diagonal form which has higher
entanglement content and Bell non-local correlations.

It should be noted that local filtering for two qubits is 
a special case of entanglement distillation~\cite{Entanglement, conc_wooters} when local operations are 
performed on the level of single copy of the quantum state. In the present 
article we restrict access of Alice and Bob to single copies and then calculate the secure key rate after local filtering.

\section{Results}
\label{sec:res}
In this section, we first develop a geometrical
representation of correlations to study the Bell-CHSH
violation and QBER for arbitrary two-qubit
states. We make the following assumptions:
\begin{enumerate}
	\item We consider only those entanglement based protocols for which 
	the secure key rate is a function of QBER only and,
	\item Bell-CHSH violation is necessary in order to distill a secure key from the 
	correlations of Alice and Bob.
\end{enumerate}

These assumptions are quite logical and are implicit in many QKD protocols including the 
CHSH protocol~\cite{Acin_sec}. Under these assumptions we apply our geometrical representation to explicitly
identify states which can provide optimal security and states which
are unusable for
a fixed Bell-CHSH violation for QKD. This geometrical
representation offers a useful visualization of two-qubit
states from a QKD perspective. Next we present a new QKD protocol which involves local filtering to improve the key rate. We conclude this section with explicit examples  of states which initially do not show Bell-CHSH violation but shows non-zero secure  key rate upon local filtering. 

\subsection{Geometrical representation of correlations}
\label{sec: geo}
As detailed in Sec.~\ref{sec: bell}, all two-qubit states
can be parameterized by the two largest singular values of
the real correlation matrix $T$ so far as the violation of
the Bell-CHSH inequality is concerned. For a
bonafide
quantum state all the singular values of the $T$ matrix must
satisfy $|\lambda_i|\leq1$
and $\sum_i \lambda_i^2\leq 3$. States lying outside this
constrained region are unphysical and do not
correspond to valid density matrices.
For the sake of simplicity we only consider the region
$0\leq\lambda_1\leq 1$ and $0\leq\lambda_2\leq 1$ as all the
arguments presented below apply equally well to the 
other valid regions.

\begin{figure}[h]
\includegraphics[scale=1]{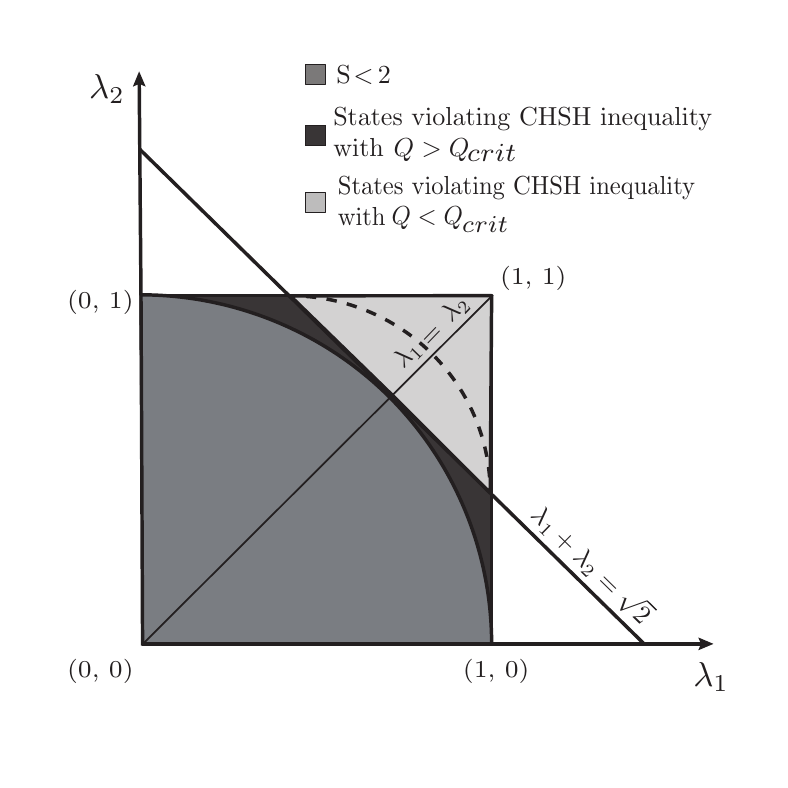}
\caption{A geometrical representation of the
Bell-CHSH inequality and the QBER $Q$ parameterized by 
$\lambda_1$ and $\lambda_2$. The dark grey region corresponds to states 
which violate the Bell-CHSH inequality but offer $Q>Q_{crit}=0.14$. These states are 
therefore unusable for QKD. Only the states lying in the light grey region offer a secure key rate
while also violating the 
Bell-CHSH inequality. }
\label{fig: violation}
\end{figure}

The geometrical representation of the two-qubit states
parameterized by the two largest singular values of the
correlation matrix $T$ is depicted in Fig.~\ref{fig: violation}.
Here all physical states are represented by shaded regions
while the unshaded region corresponds to parameter range with no corresponding
bonafide quantum state. In this representation all the states
with fixed value $S$ of the expectation
value of the CHSH operator lie on the circular arc $\lambda_1
^2 + \lambda_2^2 = S^2/4$. Therefore, all the physical
states that do not violate the Bell-CHSH inequality lie
within the disc of unit radius $\lambda_1 ^2 + \lambda_2^2
\le 1$, as can be seen from Eq.~\eqref{eq: maxs2}, while all
physical states lying outside this region show a violation.
Thus, for a given physical state its distance from the origin
quantifies the Bell-CHSH correlation and if this
distance is above $1$ the state violates the CHSH
inequality.
In this geometric representation, the QBER
$Q$of Eq.~\eqref{eq: error} is represented by straight lines with slope $-1$, i.e,
$\lambda_1 + \lambda_2 = m$  (Fig.~\ref{fig: violation}), where $m$ is the $y$-intercept.
These states offer the same $Q = \frac{1}{4}\left(2 - m\right)$. Increasing values of $m$ for
the straight lines corresponds to a decreasing QBER.

\subsection{Characterization of states based on the
geometrical representation}
\label{sec: res}
We are now ready to use the geometrical representation described above to identify states  which  violate the Bell-CHSH inequality, but cannot be used to distill a secure key rate.  This way of identifying states which are useless for QKD,  is stricter than the one identified earlier~\cite{lutkenhaus_key}. We also identify a set of
states most suitable for experimentally implementing
entanglement assisted QKD protocols with fixed violation of
the Bell-CHSH inequality from the perspective of minimum error rate.  

It is clear from  Fig.~\ref{fig: violation} that the set of
states having the same Bell-CHSH value $S$ do not share the
same error rate $Q$, hence the secure key rate
$r_{{min}}$ is also different. Considering entanglement
as an expensive resource, the variation
 in the error rate for
the same value of $S$ indicates that some states are more
suitable for performing QKD than others despite having the
same Bell non-locality. This also implies that the violation
of Bell-CHSH inequality alone cannot provide a
characterization of the security in an entanglement assisted
QKD protocol.

Note that all the classical states saturating the Bell-CHSH
bound lie on the circle $\lambda_1^2 + \lambda_2^2 = 1$.
However, as noted in Sec.~\ref{sec: geo} all these states
do not share the same error rate $Q$. The set of states
offering the least error rate $Q$ for a given value of $S$
lie on the line which is tangent to the circle of radius
$S/2$ and will satisfy $\lambda_1 = \lambda_2 =
S/2\sqrt{2}$.
Therefore the set of local states saturating the Bell-CHSH
inequality and offering the least error rate lie on the
point $\left(\frac{1}{\sqrt{2}}, \frac{1}{\sqrt{2}} \right)$
(Fig.~\ref{fig: violation}). Since Bell-CHSH violation is
necessary for the security of the QKD protocol, the states
corresponding to the point $\left(\frac{1}{\sqrt{2}},
\frac{1}{\sqrt{2}} \right)$ offer no security. We define the
error rate at this point as the critical error rate and is
given as $Q_{\text{crit}} = \frac{1}{4}\left(2-\sqrt{2}\right)
\approx 0.14$, which is the maximum allowed error for any QKD protocol to be secure, irrespective of the type of attacks by eavesdroppers. All the state on the line $\lambda_1 +
\lambda_2 = \sqrt{2}$ have the same critical error rate
(Fig.~\ref{fig: violation}).
All valid quantum states lying
below this line posses higher error rate and therefore can
not be used for secure QKD. To summarize, all the states above $\lambda_1^2+\lambda_2^2 = 1$ 
violate Bell-CHSH inequality and all the states below $\lambda_1+\lambda_2 = \sqrt{2}$ have 
QBER more than $Q_{\text{crit}}$; hence unusable for secure QKD.
The region of intersection between these two regions contain states which 
are useless for QKD, even though they  are Bell non-local.  

The critical line, $\lambda_1+\lambda_2=\sqrt{2}$ provides the theoretically maximum tolerable  error rate for carrying out secure QKD using  Bell-CHSH violation as a necessary requirement. It may happen that for a given QKD protocol $Q_{crit}$ is smaller than the one we obtained. For example, $Q_{crit}$ in the protocol presented in~\cite{lutkenhaus_key}  is $Q'_{crit} = 0.11$ which is smaller than the theoretical critical value calculated above.  This is because $Q'_{crit}$ is obtained from a particular form of $r_{min}$ which depends on the attacks chosen by the eavesdropper and communication scheme employed. 
For the huge  class of attacks detailed in Ref.~\cite{lutkenhaus_key}, $r_{min}$ takes on the form as given in Eq.~\eqref{eq: securekey}.
The value of $Q'_{crit}=0.11$ has also been shown to 
be optimal under two way communication schemes~\cite{Acin_sec,two_way_EB}.
However, there might exist attacks that may not conform to the  aforementioned form of secure key rate. 
Furthermore, any future protocols employing $n$-way communication schemes ($n>2$), 
the tolerable error rate may also be improved.
Our results are more general as we do not assume any particular form of attack by Eve or any specific 
communication scheme, but rather focus on the fact that the correlations between Alice and Bob violate the CHSH inequality.

In order to perform entanglement based QKD protocols, one needs to transmit part of quantum systems through noisy channels, which can affect the state of the quantum systems. Therefore,
it is not always possible to achieve the quantum maximum of the Bell-CHSH inequality. In those cases, it is desirable to identify states most suitable for QKD for a particular violation. These states should have the property of offering the least error rate for a fixed violation of the CHSH inequality. As detailed in Sec.~\ref{sec: geo}, these states are identified with the points $|\lambda_1| = |\lambda_2|>1/\sqrt{2}$. 

In the non-ideal QKD scenario, Alice and Bob may share
states violating the Bell-CHSH inequality but with an error
rate higher than $Q_{crit}$. It is then desirable to
transform these states such that the error rate is
reduced below the critical value
and the states can be used to distill a secure key. 
Since we were carrying out QKD between
remote locations, such transformations will have to be local
operations performed by Alice and Bob. 
However, local
operations cannot increase the violation of the Bell-CHSH
inequality unless we sacrifice some of the copies from
the ensemble~\cite{filtering_expt}. In the following subsection, we present a QKD
protocol which incorporates local-filtering operations,
to concentrate the Bell-CHSH correlations in order to
enhance the secure key rate.

\subsection{QKD protocol using local filtering operations}
\label{sec:filtering2}

\begin{figure}
	\centering
	\includegraphics[scale=0.9]{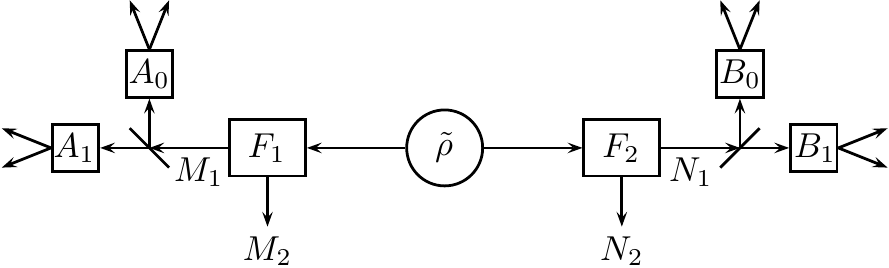}
	\caption{A schematic diagram to implement the modified QKD 
	protocol using local filtering operations.
Each party shares an initial entangled state $\tilde{\rho}$ on which they both apply local 
filters denoted by $F_1$ and $F_2$. Using classical communication, the parties discard the events
when any of the parties observed the outcome $M_2$ or $N_2$. Only when both the parties 
observe $M_1$ and $N_1$ do they proceed to perform the measurements $A_0$, $A_1$ and $B_0$, $B_1$
as dictated by the protocol.}\label{Fig:LF-QKD}
\end{figure}

In the new QKD protocol, a source is generating pairs of qubits in maximally entangled states and sending them to Alice and Bob through a channel. Due to the noisy channel and the presence of eavesdroppers, the state received by Alice and Bob is a mixed state $\tilde{\rho}$, which they can be determined by performing full state tomography before starting the QKD protocol. Hence, we start our protocol by assuming that Alice and Bob share entangled pairs of qubits in the states $\tilde{\rho}$  with Bell-CHSH value  $S$. Let $M_1$ and $N_1$ be the optimal filtering operators  for concentrating entanglement in the state $\tilde \rho$ and $M_2 = \sqrt{\mathds{1} - M_1^\dagger M_1}$ and $N_2 = \sqrt{\mathds{1} - N_1^\dagger N_1}$  (as described in \ref{sec:filtering}). $\{A_0,A_1\}$  and $\{B_0,B_1\}$ are the dichotomic observables in Alice and Bob's lab, respectively.  The modified protocol consists of the following steps:
\begin{enumerate}
\item First Alice and Bob perform local measurements using $\{M_i\}$ and $\{N_j\}$  measurement settings followed by the measurement of $\{A_0,A_1\}$  and $\{B_0,B_1\}$ on their respective subsystems. 
\item Alice and Bob announce the outcome of the measurement in $\{M_i\}$ and $\{N_j\}$ measurement  settings and the choice of the measurement operators $\{A_0,A_1\}$ and $\{B_0,B_1\}$ for each of the qubit pair.
\item They consider only the qubit pairs for which $M_1$ and $N_1$ clicked, i.e., the pairs for 
which the local filtering was successful. Then they reconcile their measurement basis $\{A_0,A_1\}$ and $\{B_0,B_1\}$ 
and discard the qubits for which the measurement was performed in different bases.
\end{enumerate}
Since, most of the implementations of QKD protocols use photons, and any measurement tends to destroy them, the local filtering followed by the measurement in $\{A_0,A_1\}$ and $\{B_0,B_1\}$ can be done by post-selection. An optical experimental setup for local filtering consists of performing a binary outcome 
POVM measurement. The measurement can be viewed as having two outcome modes
such that photon exits through one of them with the state corresponding to the 
outcome as shown in Ref.~\cite{filtering_expt, povm_implementation}. It is to be noted that
no measurement is made on the photon as it would destroy the state. For the case when the photons 
exit through the modes $M_1$ and $N_1$, Alice and Bob can then perform 
their respective measurements on their transformed photon states.  For the rest of the 
cases when filtering is unsuccessful, they discard the photon states.
In Fig.~\ref{Fig:LF-QKD} we sketch the outline of the 
modified QKD protocol using local filtering as post-selection.

 The QKD protocol presented above relies on the fact that we
can successfully filter an ensemble of two qubit partially
entangled states into a smaller ensemble with higher
entanglement.
The states with enhanced entanglement are used for QKD
while the other states are discarded. In this process one
can transform states useless for QKD
into states useful
for QKD. 
 The probability
of success in the filtering process is
$P_{\text{succ}} = \text{Tr}[(M_1\otimes N_1)\rho (M_1^\dagger \otimes
N_1^\dagger)]$.

\subsection{Example of states that do not violate Bell-CHSH inequality but can be used for QKD}

\begin{figure}
	\includegraphics[scale=1]{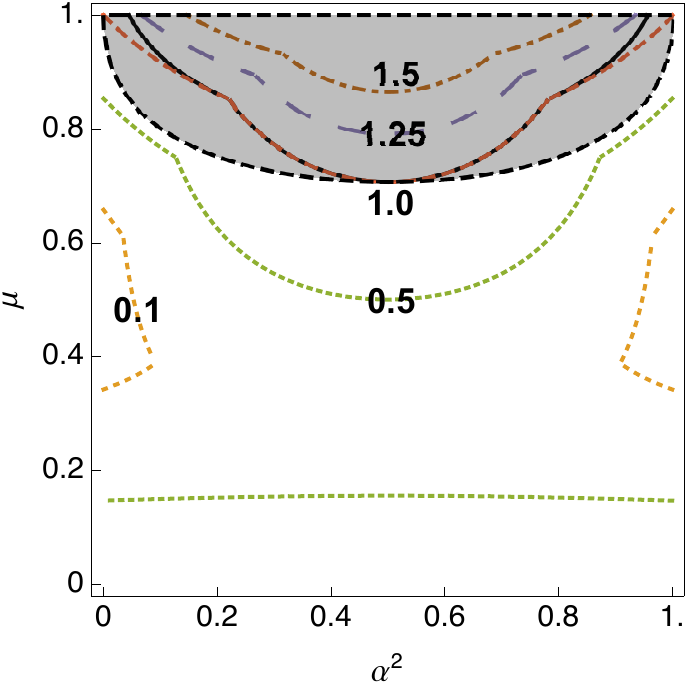}
	\caption{Contour plot of states with parameters $\alpha$ and $\lambda$ as 
	given in Eq.~\eqref{eq:example2} with varying values of $\lambda_1^2+\lambda_2^2$. In order to exhibit 
 Bell-CHSH violation it is required that $\lambda_1^2+\lambda_2^2>1$ (Red dashed). For the purpose
 of QKD it is required that $\lambda_1+\lambda_2>\sqrt{2}$ (Black solid). All states lying 
below this contour exhibit a higher error rate than $Q_{crit}$ and it can be seen that some 
of them still
exhibit Bell-CHSH violation. The set of useless states that can be made useful by local 
filtering is 
given by the region in grey.}
\label{fig:fig3}
\end{figure}

In this section we present a family of mixed states which do not show any Bell-CHSH violation; however, upon local filtering one can distill non-zero secure key rate from these states. Consider the class of states with density operators given by
\begin{equation}
\begin{aligned}
\rho=\frac{1}{4}
\left[
\mathds{1}\right.&\otimes\mathds{1} +
\mu(\alpha^2-\beta^2)(\sigma_z\otimes\mathds{1}-\mathds{1}\otimes\sigma_z)
\\
&
+ (1-2\mu)\sigma_z\otimes\sigma_z - 2\mu \alpha
\beta (\sigma_x\otimes\sigma_x + \sigma_y\otimes\sigma_y)
\left. \right],
\end{aligned}
\label{eq:example2}
\end{equation}
where $\alpha,\beta \in \mathcal{R}$, $\alpha^2+\beta^2 = 1$
and $0<\mu\leq 1$. These states have been studied extensively under local filtering 
operations \cite{Gisin1996}. The behaviour of these states for fixed Bell-CHSH violation
and error rate is plotted in Fig.~\ref{fig:fig3}. The dark grey region depicts the set of states which can be filtered to
states which can violate the Bell-CHSH inequality and have $Q<Q_{crit}$. The contour corresponding to $Q_{crit}$
characterized by $\lambda_1+\lambda_2=\sqrt{2}$ is given by the black solid line in Fig.~\ref{fig:fig3} and it is 
seen that there exist states which violate Bell-CHSH inequality and still have $Q>Q_{crit}$. However, 
these states can also be filtered.
As an example 
we consider the  state $\rho$ with $\alpha = 0.9$,  $\beta = 0.4538$ and  $\mu = 0.85$, 
which has the following properties,
\begin{equation}
\begin{aligned}
\lambda_1^2+\lambda_2^2 = 0.9347,\\
\lambda_1+\lambda_2 = 1.3669.
\end{aligned}
\end{equation}

This is an example of a state that does not violate
Bell-CHSH inequality and is therefore useless 
for QKD. 
The entanglement of formation from this state turns out to be 
\begin{equation}
\begin{aligned}
E(C(\rho)) &=h\left(\frac{1+\sqrt{1-C^2}}{2}\right)\\
&=0.3722, 
\end{aligned}
\end{equation}
where $C$ is the concurrence of the quantum state and $h(x)
= -x\log_2x-(1-x)\log_2(1-x)$ is the binary entropy. This
again states that the state $\rho$ considered
above can provide some secure key. 

After applying optimal local filtering operations as
detailed above, we get the state $\rho'$ with the 
following properties:
\begin{equation}
\begin{aligned}
\lambda'^2_1+\lambda'^2_2 = 1.3329,\\
\lambda'_1+\lambda'_2 = 1.6327.
\end{aligned}
\end{equation}
The resultant state $\rho'$ is seen to violate the Bell-CHSH
inequality with $Q<Q_{crit}$, indicating that it is now a
useful state for QKD. Consequently, the keyrate $r$ for the transformed state can be
calculated as
\begin{equation}
\begin{aligned}
r &= P_{\text{succ}} r_{min}\\
&=0.091 ~ \text{bits},
\end{aligned}
\end{equation}
where $P_{\text{succ}} = 0.799$ and $r_{min} = 0.110$ bits using Eq.~\eqref{eq: securekey}.

It is worthwhile to note that under the action of 
white noise, the state in Eq.~(\ref{eq:example2}) will be transformed as
\begin{equation}
\begin{aligned}
\tilde{\rho} &= \frac{(1-p)}{4}\mathds{1}\otimes\mathds{1}+ p \rho,
\end{aligned}
\end{equation}
which leaves the total state in a similar form as in Eq.~\eqref{eq:example2} with 
new coefficients $\mu', \alpha'$ and $\beta'$. Consulting the Fig.~\ref{fig:fig3}, we can determine if the noise added state can be made useful using local filtering or not for the new coefficients $\mu', \alpha'$ and $\beta'$.

 It should also be noted that at the
level of single-copy distillation, the local filtering
operations considered above have been shown to be optimal
for concentrating entanglement and Bell
non-locality~\cite{Filtering}.  Therefore the key rates
obtained after applying local filtering, are the best that
can be achieved, given access to individual copies only for the 
entanglement based QKD protocol.

Further, according to~\cite{Filtering}, Bell-diagonal
states cannot be filtered further. From Fig.~\ref{fig:
violation} it can be easily seen that there exist such Bell
diagonal states which exhibit Bell-CHSH violation, having
$Q>Q_{crit}$ and which  cannot be filtered. These states remain
useless for QKD even after filtering, thereby indicating
that Bell-CHSH violation is not a sufficient condition
either.

\section{Conclusion}
\label{sec: conc}
We develop a geometrical representation for two-qubit correlations to quantitatively analyse the relationship between the secure key rate of a QKD protocol and the violation of the Bell-CHSH inequality.  The usefulness of this geometrical representation is demonstrated by showing that states sharing the same non-local correlations do not necessarily share the
same secure key rate. This leads to an important conclusion
that some states are more apt for performing QKD efficiently
than others, even when they share the same non-local
correlations.

For fixed (non-maximal) Bell-CHSH violation the states
 that are optimally suited for performing QKD are identified,
which can be useful when Alice and Bob share a
non-maximally entangled state.
We use the threshold error rate requirement for
security 
to identify a class of states which cannot be
used for QKD, even though they exhibit a violation of the
Bell-CHSH inequality.
This is an improvement over a previous result and has profound
experimental implications to develop QKD protocols with non-maximal violation
of Bell-CHSH inequality. Such states which are deemed 
useless for QKD can be seen as a result of the specificity of the protocol considered or 
because of errors arising due to preparation, transmission or measurements.
To harness the entanglement present in states that do not
violate Bell-CHSH inequality we employed
local filtering
operations and found that the performance of such
states can be greatly improved in terms of providing key
rate for QKD. The local filtering operations considered is a
special subclass of entanglement distillation dealing with single copies. Under 
the paradigm of single copy distillation not all entangled states can provide a 
secure key as compared to multicopy distillation in which all 
two qubit entangled states can be used to distill some secure key. However, 
multicopy distillation is harder to achieve experimentally than local
filtering. Further, the protocol for local filtering described has been shown 
to be optimal in the case of single copies~\cite{Filtering} and the secure key 
rate obtained under these operations is the best that can be achieved.
We explicitly provided examples when
the original state exhibits Bell-CHSH violation but has
$Q>Q_{crit}$ and states which do not violate the Bell-CHSH
inequality. It is seen that in both cases local filtering
offers improvement in terms of secure key rate. Our work paves 
the way for efficient experimental realization of  QKD protocols where Bell-CHSH violation is a necessary resource.

{\bf Acknowledgements.---} J.S. would like to acknowledge
funding from UGC, India. S.~K.~G.~ acknowledges the
financial support from SERB-DST (File No. ECR/2017/002404).


%

\end{document}